\documentclass[12pt]{article}
\usepackage{amssymb,graphicx}
\usepackage{epstopdf}
\usepackage{amsmath,amsfonts}
\usepackage{epsfig}

\newcommand{\be}{\begin{equation}}
\newcommand{\ee}{\end{equation}}
\newcommand{\bea}{\begin{eqnarray}}
\newcommand{\eea}{\end{eqnarray}}
\newcommand{\bg}{\begin{gather}}
\newcommand{\eg}{\end{gather}}
\newcommand{\bseq}{\begin{subequations}}
\newcommand{\eseq}{\end{subequations}}

\renewcommand{\ln}{\mathop{\rm ln}\nolimits}

\newcommand\lp{\left(}
\newcommand\rp{\right)}

\newcommand\mg{m_{\tilde{G}}}
\newcommand\mpl{M_{\text{Pl}}}

\begin{document}
\begin{flushright}
\end{flushright}
\vspace{10pt}
\begin{center}
  {\LARGE \bf Is gravitino still\\ a warm dark matter candidate?} \\
\vspace{20pt}
D.~Gorbunov, A.~Khmelnitsky and V.~Rubakov\\
\vspace{15pt}
\textit{ Institute for Nuclear Research of the Russian Academy of Sciences,\\
 60th October Anniversary Prospect, 7a, 117312 Moscow, Russia}\\

    \end{center}
    \vspace{5pt}

\begin{abstract}
We make use of the phase space density approach to discuss gravitino
as a warm dark matter candidate. Barring fine tuning between the
reheat temperature in the Universe and superparticle masses, we find
that \emph{warm} gravitinos have both appropriate total mass density,
$\Omega_{\tilde G} = \Omega_{DM} \simeq 0.2$, and suitable primordial
phase space density at low momenta provided that their mass is in the
range $1\,\text{keV} \lesssim \mg \lesssim 15~\text{keV}$, the reheat
temperature in the Universe is low, $T_R \lesssim 10~\mbox{TeV}$, and
masses of some of the superparticles are sufficiently small, $M
\lesssim 350~\mbox{GeV}$.  The latter property implies that the
gravitino warm dark matter scenario will be either ruled out or
supported by the LHC experiments.

\end{abstract}


\section{Introduction and summary}

The predictions of the $\Lambda$CDM model are in outstanding
consistency with the bulk of cosmological observations~\cite{LCDM}
(see also Ref.~\cite{PDG} and references therein).  Yet there are
clouds above the collisionless cold dark matter scenario, which have
to do with cosmic structure at subgalactic scales. Three most notable
of them are missing satellites~\cite{satellites}, cuspy galactic
density profiles~\cite{cuspy} and too low angular momenta of spiral
galaxies~\cite{angular}.  All these suggest that CDM may be too cold,
i.e.  that the vanishing primordial velocity dispersion of dark matter
particles may be problematic. Hence, one is naturally lead to consider
warm dark matter (WDM) scenarios~\cite{BOT, Avila-Reese,
WDM,gravitino-WDM}.

There are several ways to describe the difference between WDM and CDM
scenarios.  The most prominent one is that warm particles filter
primordial power spectrum on small scales, and thus the formation of
small halos is suppressed. The filtering scale must be small enough,
since the power spectrum shows no significant deviations from the CDM
prediction on scales within reach of current observations. This leads
to constraints on the primordial velocity dispersion of WDM
particles~\cite{LyAlpha}.  On the other hand, in order to improve on
structure formation, the filtering scale must be of the order of the
scale of missing satellites, which is believed to be of order $10^7 -
10^8 M_\odot$~\cite{Mateo}.

\begin{figure}[t]
\begin{center}
\includegraphics[scale=1.2]{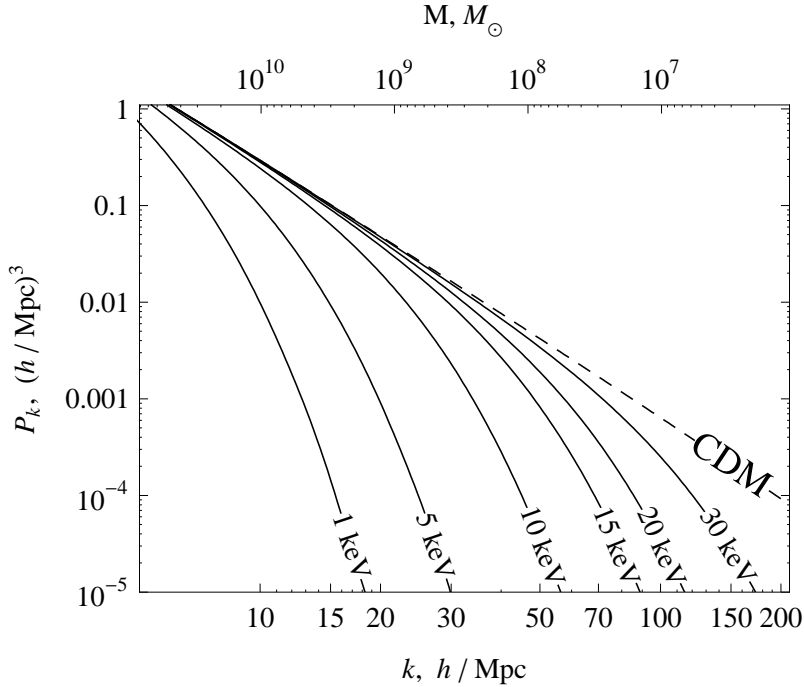}
\caption{\small{Linear matter power spectrum for standard $\Lambda$CDM
 cosmology (dashed  line) and $\Lambda$WDM (solid lines)
 assuming the distribution of WDM particles as given
 by~\eqref{eq:th-dist} with masses $m = 1, 5, 10, 15, 20 $ and 30 keV
 and $g_* = g_\text{MSSM}$.\label{fig:Pk} }}
\end{center}
\end{figure}

We have calculated linear matter power spectrum in
$\Lambda$WDM cosmology assuming that dark matter particles have the
Fermi--Dirac 
primordial distribution function, 
normalized to correct present total density:
\begin{equation}\label{eq:th-dist}
 f(p) = \frac{\rho_\text{DM}}{6 \pi\, \zeta(3)\, m\,T_{0,eff}^3}\,
\frac{1}{e^{p/T_{0,eff}}+1} \; ,
\end{equation}
where $m$ is the WDM particle mass, $T_{0,eff}\equiv T_0
\lp\frac{g_{*0}}{g_*}\rp^{1/3}$; $g_*$ and $g_{*0} \equiv
\frac{43}{11}$ are the effective number of relativistic degrees of
freedom at the epoch of dark matter particle production and at present
epoch, respectively.  To this end we have modified the Boltzmann
evolution equations implemented in the Code for Anisotropies in the
Microwave Background (\textsc{camb})~\cite{CAMB}. Figure~\ref{fig:Pk}
presents the resulting $\Lambda$WDM power spectrum for $m = 1, 5, 10,
15, 20$ and 30 keV (solid) in comparison with $\Lambda$CDM
(dashed). In the WDM case, $g_*$ is chosen to be equal to
$g_\text{MSSM} = 228.75$, the maximum number of relativistic degrees
of freedom equilibrated in plasma in the framework of MSSM. One
concludes that the power spectrum is suppressed by about an order of
magnitude on the scales corresponding to $10^8 M_\odot$ and smaller
provided the WDM particle mass is smaller than about 15~keV. Thus we
consider a particle as a WDM candidate if its mass obeys \be
\label{eq:notcold} m
\lesssim 15\,\text{keV} \; . 
\ee 
Of course, this is an indicative figure,
not a strict upper limit.

Another way to quantify the notion of warm dark matter is to make use of the
phase space density approach~\cite{Hogan, Dalcanton, Boyanovsky}.  Its key
ingredient is the ratio between the mass density and the cube of the
one-dimensional velocity dispersion in a given volume, $Q \equiv
\rho/\sigma^3$. On the one hand, this quantity is measurable in galactic halos;
on the other hand, it can be used as an estimator for coarse-grained
distribution function of halo particles. Namely, for non-relativistic dark
matter particles
\[
Q \simeq m^4\cdot \frac{n}{\langle \frac13 p^2\rangle^{3/2}} \; ,
\]
where $m$ is the mass of these particles and $n$ is their average number density
in a halo.  Assuming that the coarse-grained distribution of halo particles is
isotropic, $f_{halo}(\mathbf{p}, \mathbf{r}) = f_{halo}(p, r)$, one estimates
\[
\frac{n}{\langle p^2\rangle^{3/2}} = \frac{\left[\int{f_{halo}(\mathbf{p},
\mathbf{r}) d^3\mathbf{p}}\right]^{5/2}} {\left[\int{f_{halo}(\mathbf{p},
\mathbf{r}) \mathbf{p}^2 d^3\mathbf{p}}\right]^{3/2}} \sim f_{halo} (p_*, r) \;
,
\]
where $p_*$ is a typical momentum of the dark matter particles. In this way the
magnitude of the coarse-grained distribution function in galactic halos is
estimated as
\begin{equation}
\label{eq:Qf}
  f_{halo} \simeq \frac Q {3^{3/2}\, m^4} \; .
\end{equation}

Coarse-grained distribution function is known to decrease during
violent relaxation in collisionless systems~\cite{LyndenBell}. Hence,
the primordial phase space density of dark matter particles cannot be
lower than that observed in dark halos. This leads to the
Tremaine--Gunn-like constraints on dark matter models~\cite{Hogan,
Boyanovsky}.  The strongest among these constraints are obtained by
making use of the highest phase space densities observed in dark
halos, namely those of dwarf spheroidal galaxies
(dSph)~\cite{Mateo,Dalcanton}.  dSph's are the most dark matter
dominated compact objects, and seem to be hosted by the smallest halos
containing dark matter~\cite{Mateo}. In recently discovered objects
Coma Berenices, Leo IV and Canes Venaciti II, the value of $Q$
ranges from $5\cdot10^{-3}\,
\frac{M_{\odot}/\text{pc}^3}{\lp\text{km}/\text{s}\rp^3}$ to
$2\cdot10^{-2}\,
\frac{M_{\odot}/\text{pc}^3}{\lp\text{km}/\text{s}\rp^3}$~\cite{Simon}.
In what follows we use the first, more conservative value, 
\be Q =
5\cdot10^{-3}\,
\frac{M_{\odot}/\text{pc}^3}{\lp\text{km}/\text{s}\rp^3} \; .
\label{new3}
\ee
By requiring that the primordial distribution function 
exceeds the coarse-grained one, $f>f_{halo}$, one arrives at the
constraint
\be
3^{3/2} m^4 f > Q \; .
\label{new1}
\ee
This constraint gives rise to a reasonably well defined lower
bound on $m$ in a given model. 

If the primordial distribution is such that \eqref{new1} is barely satisfied,
the formation of high-$Q$ objects like dSph's is suppressed. 
In fact, it may be suppressed even for larger $f$, since the
coarse-grained distribution function may decrease considerably
during the evolution. The parameter
\[
\Delta \equiv \frac{3^{3/2} m^4 f}{Q}
\]
shows how strongly the coarse-grained distribution function $f$ 
must be diluted
due to relaxation processes in order that the formation of dense
compact dark matter halos be suppressed.  It 
is known from simulations that the phase space density
decreases during the structure formation. In particular, during
  the nonlinear stage it decreases  
by a factor of $10^2$
to $10^3$~\cite{Peirani}, 
or possibly higher. 
Hence, the primordial 
distribution function of WDM particles should be such that $\Delta
\gtrsim 10^2 - 10^{3}$. At least naively, obtaining the dilution factor
in  a given model in the ballpark $\Delta = 1 - 10^3$ would indicate
that the primordial phase space density is just right to make
dwarf galaxies but not even more compact objects. Interestingly, we
will find that $\Delta$ is indeed in this ballpark for WDM gravitinos obeying
\eqref{eq:notcold}.

As discussed in Ref.~\cite{Madsen}, only a fraction of dark matter
particles should definitely have high phase space density
obeying~(\ref{new1}).  This fraction $\nu$ is estimated as the
relative contribution of dSph's into the total mass density of dark
matter.  Using the dwarf number density $n_{dwarf} \simeq
7\cdot10^{-2} ~\text{Mpc}^{-3}$ from~Ref.~\cite{Loveday} and assuming
the average dSph mass of order $10^7 M_{\odot}$~\cite{Mateo}, one
estimates 
\be \nu \simeq
\frac{\Omega_{\text{dSph}}}{\Omega_{\text{DM}}} \simeq 10^{-5} \; .
\label{new2}
\ee 
To be on conservative side, we impose the constraint \eqref{new1}
on this fraction of WDM particles only. Also, we 
calculate the value of 
$\Delta$ for this fraction.
One expects that once the right fraction of the dark matter particles has
the high phase space density, the most compact objects are produced in
right numbers (and not overproduced). We have checked that our results would
change very little if we used an estimate for $\nu$ differing from~(\ref{new2})
even by an order of magnitude, i.e. $\nu = 10^{-6} - 10^{-4}$.

To summarize, we consider a dark matter model viable if a fraction~(\ref{new2})
of its particles has primordial phase space density
obeying~(\ref{new1}) with $Q$ given
by Eq.~(\ref{new3}).

In this paper 
we make use of this phase space density criterion 
together with the bound \eqref{eq:notcold}
to examine light
gravitino as a warm dark matter candidate, assuming that R-parity is conserved
and hence gravitino is stable.  We find that gravitino 
mass should be in the range
\[
1\,\text{keV} \lesssim \mg \lesssim 15\,\text{keV} \; ,
\]
cf.~\cite{gravitino-WDM}. 

In the early Universe, light gravitinos are produced in decays of superparticles
and in scattering
processes~\cite{gravitino-production,Bolz,Pradler,Rychkov}. For so light
gravitinos, their production in decays of superparticles plays an important
role~\cite{Moroi}. We consider this mechanism in Sec.~\ref{sec:decays}, where we
also evaluate the spectrum of produced gravitinos.  In Sec.~\ref{sec:scattering}
we discuss gravitino production in scattering processes. The latter mechanism
operates most efficiently at the highest possible temperatures in the early
Universe, so the requirement that gravitinos are not overproduced restricts
severely the reheat temperature $T_R$, cf.~\cite{Moroi, reheat-bounds}; we find
that $T_R$ must be at most in the TeV range.

Most notably, gravitinos serve as warm dark matter candidates
only if other superparticles are rather light.  We find that
superparticles whose mass $M$ is below the reheat temperature should
obey \be M \lesssim 350~\mbox{GeV} \; ,
\label{new5}
\ee otherwise gravitinos are overproduced in their decays and in scattering
and/or relic gravitinos are too cold.  Barring fine tuning between the reheat
temperature in the Universe and superparticle masses, this means that gravitino
as warm dark matter candidate will soon be either ruled out or supported by the
LHC experiments.

The bound (\ref{new5}) is to be compared to the experimental bounds on
masses of gluino and quarks of the 1st and 2nd generations,
$M_{\tilde{q}, \tilde{g}} \geq 250 - 325$~GeV~\cite{PDG}.  Given the
narrow interval between these bounds, we find it disfavored that
squarks and gluinos participate in gravitino production
processes. Hence, we elaborate also on a scenario with relatively
light colorless superparticles whose masses $M$ obey~(\ref{new5}),
heavy squarks and gluinos, and reheat temperature in between, \be M
\lesssim T_{R} \ll M_{\tilde{q}, \tilde{g}} \; .
\label{case-of}
\ee In this scenario, squarks and gluinos do not play any role in gravitino
production, while the important production processes are decays and collisions
of sleptons, charginos and neutralinos.  We find that in this case, the overall
picture is consistent in rather wide range of parameters, with the reheat
temperature extending up to 10~TeV.

It is worth noting that for light gravitino we consider in this paper, the
lifetime of next-to-lightest superparticles (NLSP) 
is short, $\tau_{NLSP} \lesssim 2
\cdot \,10^{-5}\,\text{s}$.  Thus, their decays after decoupling are not
hazardous for BBN. On the other hand, in the mass range of gravitino and
superpartners we have found favored, one has $\tau_{NLSP} \gtrsim 5 \cdot
10^{-7}\,\text{s}$ and thus the NLSP decay length 
(neglecting $\gamma$-factor)
is in the range
$$
160\,\text{m} \lesssim c\tau \lesssim 7\,\text{km} \; .
$$ So, with gravitino WDM, it is likely that NLSP (if chargeless and colorless)
will freely travel through the LHC detectors.

\section{Gravitino production mechanisms}

\subsection{Production in decays}
\label{sec:decays}
We begin with the study of the light gravitino production in two-body decays of
thermalized superparticles, assuming that the reheat temperature in the Universe
exceeds considerably the masses of these superparticles.  Let us first find the
distribution function of gravitinos produced in decays of one kind of
superparticles with mass $M$.  At time $t$, these superparticles have thermal
distribution function $f_{th}(p,t)$.

The Boltzmann equation for the gravitino distribution function $f(p,t)$ is
\begin{equation*}
  \frac{\partial f(p, t)}{\partial t} - H(t) p \frac{\partial f(p, t)}{\partial
  p} = I,
\end{equation*}
where $H(t)$ is the Hubble parameter. In this section we consider the
contribution into the collision term $I$ that comes from two-body decays and is
given by
\begin{equation*}
  I = \frac1{2 |\mathbf{p}|} \int
\frac{d^3 P}{2 E (2\pi)^3} \frac{d^3 p'}{2 |\mathbf{p'}| (2\pi)^3}
(2\pi)^4 \delta^{(4)}(P - p - p')
  f_{th}(P, t) |\mathcal{M}|^2.
\end{equation*}
Here $\mathbf{P},\, \mathbf{p},\, \mathbf{p'}$ are the 3-momenta of the decaying
superparticle, gravitino and another decay product (SM particle), respectively;
$E = \sqrt{M^2 + \mathbf{P}^2}$ is the energy of the decaying particle. The
amplitude $\mathcal{M}$ is related to the decay rate in the rest frame of the
decaying particle as $|\mathcal{M}|^2 = 16\pi M \Gamma$. Neglecting the masses
of particles in the final state, one has\footnote{This formula, generally
speaking, does not work in the Higgs--higgsino sector, and in some region of the
parameter space the corresponding decays are suppressed. We treat higgsinos on
equal footing with other charginos and neutralinos in what follows.  Refining
this approximation would not change our results considerably.}~\cite{Nilles}
\begin{equation*}
\Gamma = \frac{M^5}{6\mg^2 \mpl^2} \; .
\end{equation*}
We note that both decays and scattering processes produce longitudinal gravitino
(goldstino), so the number of gravitino helicity states effectively equals
two. We also note that in the parameter range of interest, the inverse $2 \to 1$
processes, leading to disappearance of gravitino, have negligible rates.

Upon integrating over the momentum of the SM particle and over the direction of
${\bf P}$, the collision term takes the following form,
\begin{equation*}
  I = \frac{M \Gamma}{p^2} \int\limits_{E_{min}}^\infty f_{th}(P, t)\,dE \; ,
\end{equation*}
where
\[
E_{min} = p + \frac{M^2}{4p}
\]
is the minimum energy of the decaying particle capable of producing gravitino of
momentum $p$.  Of particular interest for what follows is the low momentum
region, $p \ll M,\,T$. In that case $E_{min} \gg M$, i.e., slow gravitinos are
born in peculiar decays of fast moving superparticles, which produce gravitinos
in a narrow backward cone. For this reason, the efficient production of slow
gravitinos occurs at temperatures $T \gtrsim M^2/p \gg M$.  As we will see
shortly, for relatively low reheat temperatures $T_R$ this results in a
non-trivial shape of the gravitino spectrum at low momenta, with a cutoff at
$p/T \sim M^2/T_R^2$.

It is convenient to take comoving momentum $q = a(t) \, p$ as the argument of
the gravitino distribution function. Here $a(t)$ is the scale factor, whose
present value is normalized to unity, $a(t_0) = 1$. The Boltzmann equation takes
the form
\begin{equation*}
  \frac{d f(q, t)}{d t} =
\frac{M \Gamma}{q^2} a^2(t) \int\limits_{E_{min}}^\infty f_{th}(P, t)\,dE \; .
\end{equation*}
It can be easily integrated, giving
\begin{equation*}
  f(q, t) = \int\limits_{t_R}^t~dt^\prime \frac{M \Gamma}{q^2}
a^2(t^\prime) \int\limits_{E_{min}}^\infty f_{th}(P, t^\prime)\,dE \; ,
\end{equation*}
where $t_R$ refers to the beginning of the thermal phase of the cosmological
evolution after reheating. Hereafter we assume that the production of gravitinos
is negligible at the reheating epoch.  This is of course an arbitrary assumption
reflecting our ignorance of the reheating mechanism; we expect that the
gravitino production at reheating, if any, would make the regions of favored
gravitino and superparticle masses even narrower as compared to the regions
presented below.

Since the thermal distribution function $f_{th}(P, t)$ of the decaying particles
depends on the ratio $E/T(t)$ only, it is convenient to trade the integration
over production time for the integration over temperature. To this end we use
the entropy conservation and the relation $T = \sqrt{\frac{\mpl^*}{2t}}$ with
$\mpl^* \equiv \mpl\sqrt{\frac{90}{8\pi^3 g_*}}$, valid at the radiation
domination epoch.  Thus, if the gravitino distribution function had not been
distorted by structure formation, at the present epoch it would have been given
by
\begin{equation*}
  f(q, t_0) = \int\limits_{0}^{T_R}~dT~
\frac{M \Gamma \mpl^* T_{0,eff}^2}{q^2 T^5} \int\limits_{E_{min}}^\infty
  f_{th}\left(\frac{E}{ T} \right) \,dE \; .
\end{equation*}
Here $T_{0,eff}\equiv T_0 \lp\frac{g_{*0}}{g_*}\rp^{1/3}$; $g_*$ and $g_{*0}
\equiv \frac{43}{11}$ are the effective number of relativistic degrees of
freedom at gravitino production and at present epoch, respectively; in the
framework of MSSM with all superparticles relativistic in the plasma $g_* =
g_\text{MSSM} = 228.75$ and $T_{0,eff} \approx 0.7$~K.

\begin{figure}
\includegraphics[scale=.98]{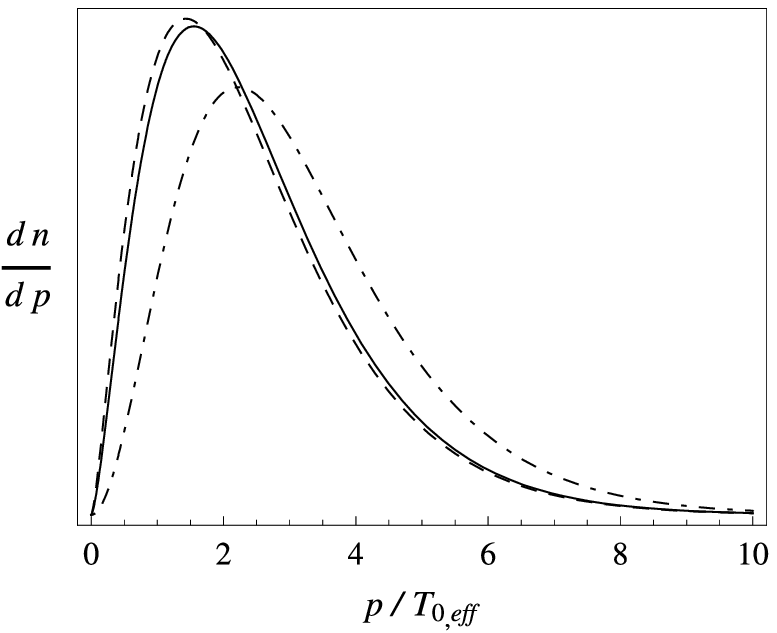}
\includegraphics[scale=.98]{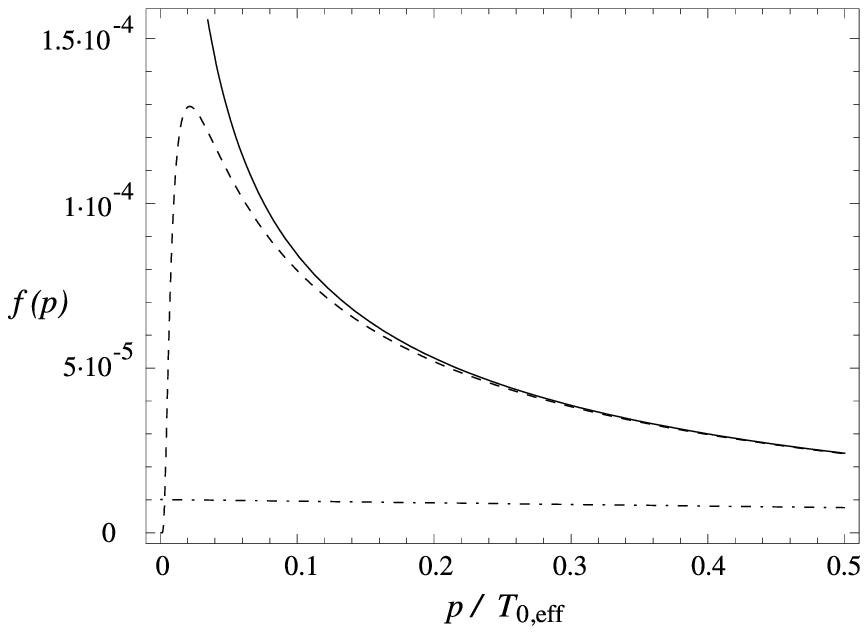}
\caption{\small{\emph{Left:} Spectra of gravitinos produced in decays of
thermalized fermions (solid) and bosons (dashed) in comparison with the
Fermi-Dirac (dash-dotted) spectrum at temperature $T_{0,\, eff}$, normalized to
the same total number of particles.  \emph{Right:} Low momentum part of the
distribution functions of gravitinos with $\mg = 10~\text{keV}$ produced in the
decays of bosons with $M = 200~\text{GeV}$ for $T_R \to \infty$ (solid line) and
$T_R = 4\,M$ (dashed line).  Dash-dotted line: Fermi-Dirac distribution at
temperature $T_{0,\, eff}$, normalized to the same total number of
particles.\label{fig:distribution} }}
\end{figure}

Changing the variables $(T, E) \rightarrow \left( z = \frac E T, x = \frac M T
\right)$ and performing the integration over $x$ we obtain finally the following
result for the primordial distribution function expressed in terms of the
momenta redshifted to the present epoch,
\begin{eqnarray}
f(p) \equiv f(q, t_0) &=& \frac83
\frac{\mpl^* \Gamma}{M^2}
\lp\frac{T_{0,eff}}{p}\rp^2 \cdot
I \left( \frac{p}{T_{0,eff}}, \frac{M}{T_R}\right)
 \nonumber \\ &=& \frac{2\sqrt5}{3 \pi^{3/2}\sqrt{g_*}}
\frac{M^3}{\mg^2\mpl} \lp\frac{T_{0,eff}}{p}\rp^2 \cdot
I \left( \frac{p}{T_{0,eff}}, \frac{M}{T_R}\right) \; ,
\label{eq:distribution}
\end{eqnarray}
where
\be
I \left( \frac{p}{T_{0,eff}}, \frac{M}{T_R}\right)
\equiv \int\limits_{z_{min}}^\infty
\left[ \lp\frac{p}{T_{0,eff}}\rp^{3/2} \lp z
-\frac{p}{T_{0,eff}}\rp^{3/2} - \lp\frac{M}{2 T_R}\rp^3 \right]
f_{th}(z)\,dz
\label{distI}
\ee
with
\[
   z_{min} = \frac{p}{T_{0 \, eff}} + \frac{M^2}{4\,T_\text{R}^2 } \frac{T_{0 \,
   eff}}{p} \; .
\]
The corresponding spectrum $\frac{dn}{dp} = 4\pi p^2 f(p)$ for $T_R \gg M$ is
shown in the left panel of Fig.~\ref{fig:distribution} in comparison with the
thermal spectrum at temperature $T_{0,eff}$ and the same total number of
particles. It is seen that gravitinos produced in decays have lower average
momentum and in this sense are cooler than thermal ones.  The overall shape of
the spectrum is not of particular interest for our purposes, however: the
formation of compact objects like dSph's depends on the low-momentum part of the
spectrum, where the phase space density $f(p)$ is high.

As we alluded to above, the low-momentum part of the spectrum depends in a
peculiar way on the ratio of the mass of the decaying particle $M$ to the reheat
temperature $T_R$. This comes out from Eq.~(\ref{distI}), in particular,
through the lower limit of integration $z_{min}$. For $T_R \to \infty$, the
distribution function at low momenta, $p \ll T_{0,eff}$, is given by
\begin{eqnarray}
f (p) &=& \frac{2\sqrt5}{3 \pi^{3/2}\sqrt{g_*}}
\frac{M^3}{\mg^2\mpl}\lp\frac{T_{0,eff}}{p}\rp^{1/2}
\int\limits_{0}^\infty z^{3/2} f_{th}(z)\,dz
\nonumber \\
&=&\frac{\sqrt{5}\zeta(5/2)}{16\pi^4\sqrt{g_*}} \frac{ M^3}{\mg^2 \mpl}
 g c_{dec}
\lp\frac{T_{0,eff}}{p}\rp^{1/2} \; .
\nonumber
\end{eqnarray}
where $g$ is the number of helicity states of the decaying particle, and
$c_{dec} =1$ for bosons and $c_{dec} = \lp 1-\frac{1}{2\sqrt{2}}\rp$ for
fermions. As expected, the combination $\mg^4 f$ entering (\ref{new1}) increases
with the gravitino mass, $\mg^4 f \propto \mg^2$, so that only light gravitinos
are warm.  We note in passing that the distribution function derived in this way
is unbounded as $p \to 0$.  However, the correct distribution function of
gravitinos --- fermions with effectively two helicity states --- cannot exceed
the value $2/(2\pi)^3$ because of Pauli-blocking. To take this into account we
simply cut the distribution function at $2/(2\pi)^3$ wherever the calculated
distribution function exceeds this value. In fact, this procedure is used almost
nowhere in the parameter space we consider in this paper, as the calculated
distribution function almost never exceeds $2/(2\pi)^3$.

For finite $T_R$, but still $T_R \gtrsim M$, the distribution
function~(\ref{eq:distribution}) no longer peaks at $p \to 0$. Instead, it has a
rather broad peak at $p/T \sim M^2/T_R^2$ and exponentially decays towards $p
\to 0$.  This is shown in the right panel of Fig.~\ref{fig:distribution}, where
we also compare the distribution function of gravitinos produced in decays with
the thermal distribution function at temperature $T_{0, \, eff}$ normalized to
the same total number of produced particles. We again see that the gravitinos
produced in decays are substantially cooler than fermions with thermal
distribution, as the maximum phase space density is substantially higher in the
former case.

It is clear from~(\ref{eq:distribution}) that the largest contribution into the
gravitino production comes from the heaviest superparticles that have ever been
relativistic in cosmic plasma. To get an idea of numerics, let us consider the
case in which $g_b$ bosonic and $g_f$ fermionic superparticle degrees of freedom
have one and the same mass $M$, and the reheat temperature is substantially
higher than $M$. Then the present number density of gravitinos produced in
decays of these superparticles is
 \[
n_0^\text{dec} = \int f(p)\,d^3 p
= \frac{3\sqrt{5}\zeta(5)}{16\pi^{5/2}\sqrt{g_*}} T_{0,eff}^3 \frac{M^3}{\mg^2
\mpl} \lp g_b + \frac{15}{16} g_f\rp \; ,
\]
and the present mass density of these gravitinos is given by
\begin{equation}
\Omega_{\tilde{G}}^\text{dec}
= \frac{\mg\ n_0^\text{dec}}{\rho_c} \approx 8\cdot10^{-4} \lp g_b
+ \frac{15}{16} g_f\rp
\lp\frac{g_\text{MSSM}}{g_*}\rp^{3/2}
\lp\frac{1\,\text{keV}}{\mg}\rp \lp\frac{M}{100\,\text{GeV}}\rp^3 \; .
\label{om-decays}
\end{equation}
A crude estimate for the gravitino mass is obtained by assuming that the
distribution function of $\nu = 10^{-5}$ of gravitinos is roughly comparable to
the Pauli-blocking value, $f=2/(2\pi)^3$.  Then the condition~(\ref{new1})
corresponds to 
\be m_{\tilde{G}} > 1~\mbox{keV} \; .
\label{mg-estimate1}
\ee 
As an example, if the heaviest superparticles are squarks of the
1st and 2nd generations and gluinos, as motivated by mSUGRA, if they
have the same mass and the reheat temperature is high enough so that
these particles were relativistic in the cosmic plasma, then $g_b =
g_{\tilde q} = 4 \cdot 3 \cdot 4 = 48$, $g_f = g_{\tilde g} = 2 \cdot
8 = 16$ and $g_* = g_\text{MSSM}$. Making use of the
estimate~(\ref{om-decays}) and the upper limit on warm
  gravitino mass \eqref{eq:notcold}, we find in this example that the common
mass of squarks and gluinos must be rather small, $M_{\tilde{q},
\tilde{g}} \lesssim 350$~GeV, 
otherwise gravitinos are overproduced. We will refine these estimates
in Sec.~\ref{sec:results}.

\subsection{Gravitino production in scattering}
\label{sec:scattering}

Gravitino production in scattering processes has been worked out in
Refs.~\cite{Bolz, Pradler} using the Braaten-Yuan prescription and hard loop
resummation. It has been reconsidered recently in Ref.~\cite{Rychkov} with the
results substantially different from those of Refs.~\cite{Bolz, Pradler} in some
regions of parameter space. We will use the approach of Refs.~\cite{Bolz,
  Pradler} with understanding that there is considerable uncertainty both in
gravitino production rate and in their spectrum, especially at relatively low
temperatures, $T \sim M$. We will further comment on this uncertainty in
Section~\ref{sec:results}.

The contribution of scattering into the gravitino production is dominated by the
processes involving the heaviest superparticles which have ever been
relativistic in the cosmic plasma. Furthermore, this contribution strongly
depends on whether or not these superparticles are colored.  In what follows we
consider two scenarios which we think are representative for realistic
supersymmetric extensions of the Standard Model. Our analysis below is
straightforwardly redone for the general case, but given the unknown
superparticle spectrum and the uncertainty in~(\ref{new1}), considering these
simple scenarios will be sufficient for our purposes.  The first scenario has
been described in the end of Sec.~\ref{sec:decays}: in this scenario the
heaviest are squarks of the 1st and 2nd generations and gluinos, and we assume
that they all have the same mass $M$ and that the reheat temperature exceeds
$M$. Given the experimental bounds, $M \geq 250 - 325$~GeV, it is clear already
from the preliminary discussion in the end of Sec.~\ref{sec:decays} that this
scenario may be consistent only in a rather narrow range of the parameter space.
Hence, we discuss also the second scenario, which is defined by the
relation~(\ref{case-of}) where $M$ is the common mass of sleptons, charginos and
neutralinos.  In the second scenario, squarks and gluinos play no role in the
gravitino production in the early Universe.  Given the strong dependence on the
mass $M$, varying the rest of SUSY parameters in either scenario does not lead
to significant changes of our results.

For the first, squark-gluino scenario, the results of Ref.~\cite{Pradler} apply
directly, so the mass density of gravitinos produced in scattering is given by
\be \Omega_{\tilde{G}}^\text{sc} \approx \omega_s {g_s}^2
\ln\lp\frac{k_s}{g_s}\rp \lp\frac{M}{100\,\text{GeV}}\rp^2
\lp\frac{1\,\text{keV}}{\mg}\rp \lp\frac{T_\text{R}}{1\,\text{TeV}}\rp \; ,
\label{strong-sc}
\ee where $g_s$ is the strong coupling constant at the energy scale $T_R$, and
$\omega_s=0.732$, $k_s=1.271$.

For the second, color-singlet scenario, the results of
Ref.~\cite{Pradler} have to be modified. To this end, we consider
electroweak scattering processes only and omit the contributions of
reactions with external squarks.  Also, we omit the squark
contributions into the thermal masses of the gauge bosons. The overall
gravitino production cross section depends on thermal masses
$m_\text{th}$ as $\ln\lp T/m_\text{th}\rp$ and thus grows as the
thermal mass decreases.  As a result, the gravitino production
cross section in our scenario is nearly 80\% of the electroweak part
obtained in Ref.~\cite{Pradler}, although $1/3$ of all processes are
omitted.  Using the modified cross sections and $g_* = 142.75$, we
find for the present mass density of gravitinos produced in $2 \to 2$
processes in primordial plasma:
\begin{equation}\label{eq:Omegasc}
\Omega_{\tilde{G}}^\text{sc} \approx \sum_{\alpha=1}^2 \omega_\alpha
{g_\alpha}^2 \ln\lp\frac{k_\alpha}{g_\alpha}\rp
\lp\frac{M}{100\,\text{GeV}}\rp^2 \lp\frac{1\,\text{keV}}{\mg}\rp
\lp\frac{T_\text{R}}{1\text{TeV}}\rp,
\end{equation}
with modified constant factors $\omega_\alpha \approx (0.152, 0.372)$
and scales in logarithms $k_\alpha \approx (1.52, 1.52)$.  Here
$\alpha=1$ and $\alpha=2$ refer to the gauge groups $U(1)_Y$ and
$SU(2)_L$, respectively, with the gauge couplings $g_\alpha = ( g',
g)$.

The estimates~\eqref{strong-sc} and~\eqref{eq:Omegasc} have considerable
uncertainties related to infrared problems existing in field theory at finite
temperature. These will translate into uncertainties in our estimates presented
in Section~\ref{sec:results}.

\section{Results}
\label{sec:results}

There are three criteria the WDM model with light gravitino should
satisfy.  First, as discussed in Introduction, gravitino would serve
as \emph{warm} dark matter provided its mass satisfies the upper
bound~(\ref{eq:notcold}), $\mg \lesssim 15\;\text{keV} $.  Another
criterion is that the present gravitino mass density should be equal
to the observed dark matter density.  In both scenarios of
Sec.~\ref{sec:scattering}, the total gravitino mass density is the sum
of contributions due to the decay and scattering processes, so that
one requires $\Omega_{\tilde{G}}^\text{dec} +
\Omega_{\tilde{G}}^\text{sc} = \Omega_{\text{DM}} \approx 0.2$. This
requirement gives one relation between the three parameters, the
masses $m_{\tilde G}$, $M$ and reheat temperature $T_R$ in each
scenario.  For the first, squark-gluino scenario we make use
of~(\ref{strong-sc}) as well as (\ref{om-decays}) with $g_b = 48$,
$g_f = 16$ and $g_* = g_\text{MSSM}$. For the second, color-singlet
scenario the appropriate expressions are~(\ref{eq:Omegasc})
and~(\ref{om-decays}) with $g_b = g_{\tilde l} = 3 \cdot (4 + 2)= 18$,
$g_f = g_{\tilde \chi} = 4 \cdot 2 + 2 \cdot 4= 16$ and $g_* =
142.75$.  In either case, scanning the reheat temperature from $T_R
\sim M$ upwards, we observe from
Eqs.~(\ref{om-decays}),~(\ref{strong-sc}) and~(\ref{eq:Omegasc}) that
this criterion gives a lower bound on the gravitino mass for given
$M$.

The third criterion is discussed in Sec.~1: about $10^{-5}$ of
gravitinos should have the primordial phase space density obeying
(\ref{new1}) with $m \equiv m_{\tilde G}$.  This criterion gives a
lower bound on the gravitino mass for given $M$ and $T_R$. This bound
has to do with the magnitude of the gravitino distribution function at
low momenta where this function is large.  Instead of calculating the
low momentum part of the distribution function of gravitinos produced
in the scattering processes, we first use the lower bound on the
overall distribution function, which is obtained by neglecting
altogether the contribution of the scattering processes into the
distribution function in the low momentum region.  The lower bounds on
$m_{\tilde G}$ obtained within this decay dominance approximation are
overestimated in comparison with those one would obtain by the
complete treatment.  To get an idea of the uncertainty introduced by
approximating the low momentum part of the distribution function by
the contribution of the decay processes only, we then add the
contribution from scattering assuming that gravitinos produced in the
latter way have thermal-shaped distribution~(\ref{eq:th-dist}), but
normalized to the total mass density, Eqs.~(\ref{strong-sc}) and
(\ref{eq:Omegasc}) in the first and second scenario, respectively,
\begin{equation}
  f^\text{sc}(p) =
\frac{\rho_c \Omega_{\tilde{G}}^\text{sc}}{6 \pi \zeta(3) \mg\,T_{0,eff}^3}\,
\frac{1}{e^{p/T_{0,eff}}+1} \; .
\label{eq:sc-dist}
\end{equation}
Within either approximation, for each set of parameters we find the
value $f$ of the phase space density, such that $10^{-5}$ of
gravitinos have the distribution function exceeding $f$, and require
that $f$ obeys~(\ref{new1}) at an allowed point in the parameter
space.

\begin{figure}[!t]
\begin{center}
  \includegraphics[scale=1]{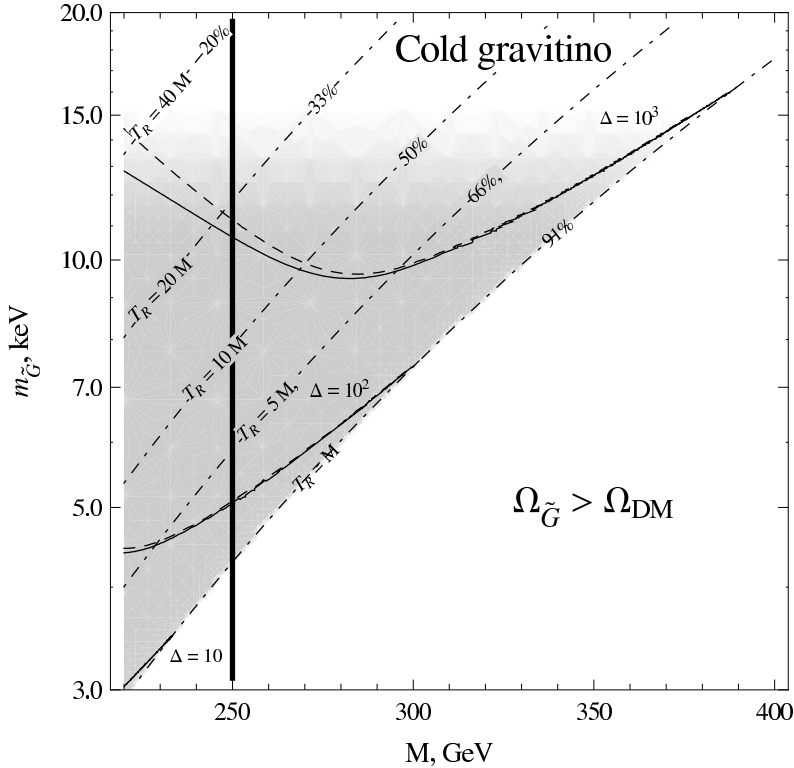}
\caption{\small{Allowed region of masses (shaded) in a scenario with heavy
gluinos and quarks of the 1st and 2nd generations, $M_{\tilde q} = M_{\tilde g}
= M$, and $T_R~\gtrsim~M$.  Contours of equal dilution factor $\Delta$ are also
shown (solid and dashed lines). The dashed lines correspond to the decay
dominance approximation, while the solid lines are obtained under the assumption
that the scattering contribution to the low momentum part of the distribution
function has the form (\ref{eq:sc-dist}).  Contours of equal $T_R/M$ are shown
with dash-dotted lines, on which the fraction of gravitinos produced in decays
is also indicated.  Conservative experimental lower bound on masses of gluinos
and squarks of the 1st and 2nd generations is indicated by solid vertical line.
}\label{fig:squark-gluino}}
\end{center}
\end{figure}

\begin{figure}[t]
\begin{center}
\includegraphics[scale=1]{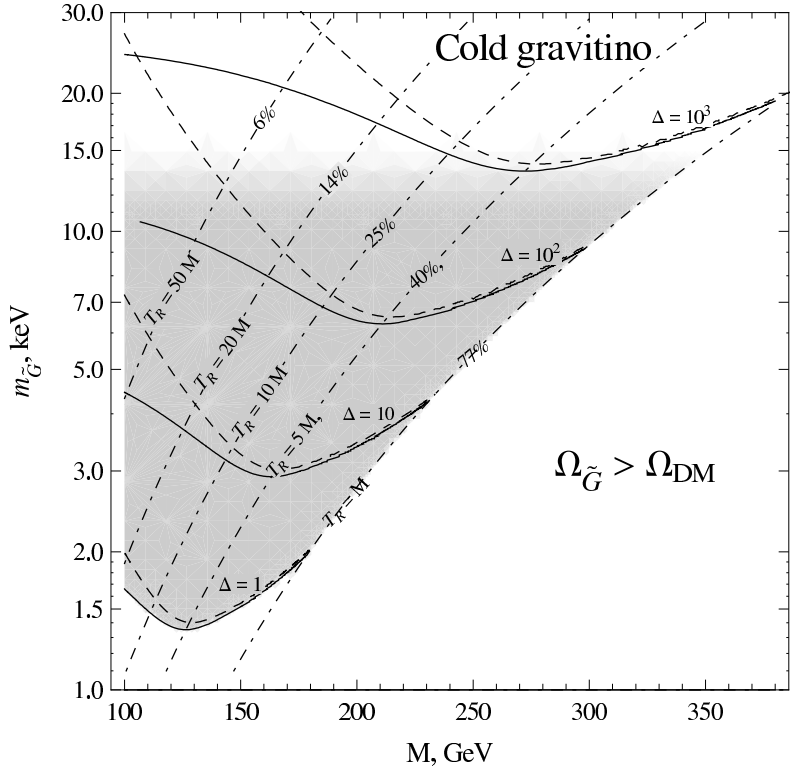}
\caption{\small{Same as in Fig.~\ref{fig:squark-gluino}, but for a
scenario with color singlet superparticles of equal mass $M$, heavy
squarks and gluinos, and intermediate reheat temperature, $M \lesssim
T_R \ll M_{\tilde{q}, \, \tilde{g}}$.}
\label{fig:color-singlet}}
\end{center}
\end{figure}

\begin{figure}[t]
\begin{center}
\includegraphics[scale = 1]{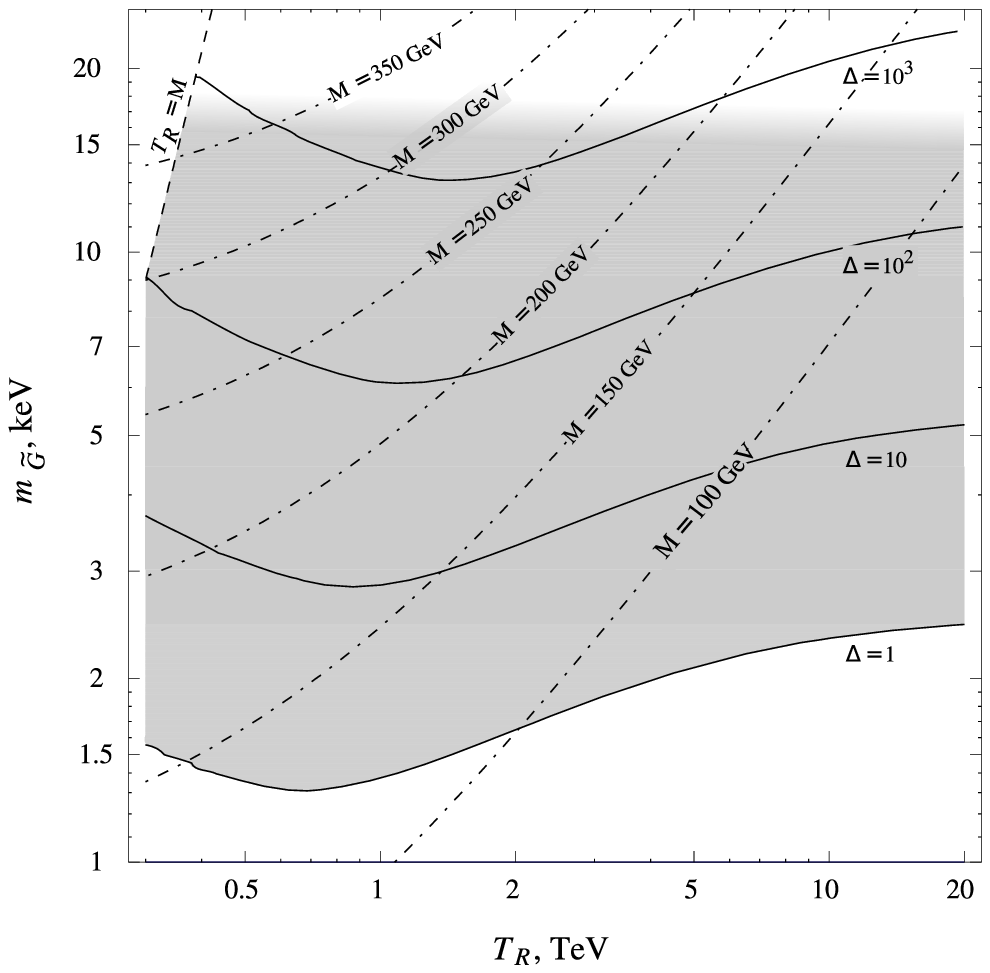}
\caption{\small{Region with gravitino WDM in the $(T_\text{R}, \mg)$
plane for the same scenario as in Fig.~\ref{fig:color-singlet}.
Contours of equal $M$ are shown with dash-dotted lines. }
\label{fig:Tm}}
\end{center}
\end{figure}

The resulting bounds in $(M, \mg)$ plane are shown in
Fig.~\ref{fig:squark-gluino} and in Fig.~\ref{fig:color-singlet} for the first
and second scenario, respectively.  Contours of equal $T_R/M$ are plotted with
dashed-dotted lines.  At the same time, these contours correspond to constant
fractions of gravitinos produced in scattering and decay channels, providing
together the correct present mass density of dark matter, $\Omega_{\tilde
G}=\Omega_{DM}$. The labels show the fraction of gravitino produced in
decays. The shaded regions are allowed by both \eqref{eq:notcold} and
\eqref{new1}.  We also show the lines of equal dilution factor $\Delta$; the
dashed lines correspond to the decay dominance approximation, while the solid
lines are obtained under the assumption that the scattering contribution to the
low momentum part of the distribution function has the form
(\ref{eq:sc-dist}).

In view of substantial uncertainty in the production of gravitinos in
scattering, the estimates for the reheat temperature should be considered as
indicative only. This is particularly relevant for the upper left parts of
Figs.~\ref{fig:squark-gluino} and~\ref{fig:color-singlet}, where production in
scattering dominates over production in decays. Also, the estimates for the
dilution factor $\Delta$ are uncertain in these parts of the parameter space,
due to the large uncertainty in the low momentum part of the spectrum of
gravitinos produced in scattering. This is reflected by the fact that dashed and
solid lines deviate significantly from each other in the upper left parts of
Figs.~\ref{fig:squark-gluino} and~\ref{fig:color-singlet}.  Furthermore, even
though the thermal-shaped distribution~\eqref{eq:sc-dist} is a plausible
approximation, we cannot exclude the possibility that scattering contribution to
the distribution function of gravitinos is much larger at low momenta as
compared to~\eqref{eq:sc-dist}. In the latter case the lines of equal dilution
factor $\Delta$ would shift even further down. In any case, the most
conservative lower bound on the gravitino mass independent of the distribution
function is given by~\eqref{mg-estimate1}.

Irrespectively of these uncertainties, we see that in both scenarios, the
relevant superparticle masses must be rather 
low, $M < 320 - 350$~GeV, provided that the reheat temperature is $T_R \gtrsim
M$.  Extending the mass range of superparticles towards larger $M$ in either
scenario would require increasingly strong fine tuning between the reheat
temperature and these masses. This fine tuning is needed to ensure that
superparticles are non-relativistic and hence not so numerous, but have just
right abundance at the beginning of the thermal stage of the cosmological
evolution to produce just right number of gravitinos.  We consider this
possibility implausible.

Figure~\ref{fig:Tm} shows the same bounds as in Fig.~\ref{fig:color-singlet} in
$(T_\text{R}, \mg)$ plane. On dash-dotted lines the total density of gravitinos
produced in both channels is equal to the observed dark matter density for
indicated superpartner masses $M$.

We conclude that unlike in the WIMP case, gravitino WDM does not
automatically have the present mass density in the right ballpark.  If
the heaviest superparticles are squarks and gluinos, and they were
relativistic in the cosmic plasma (the first scenario), the allowed
range of parameters is rather narrow, as seen from
Fig.~\ref{fig:squark-gluino}.  We consider least contrived the
possibility that the masses of sleptons, charginos and neutralinos are
in the range $M = 150 - 300$~GeV, the reheat temperature is $T_R =
200~\mbox{GeV} - 10$~TeV and the masses of gluinos and squarks are
higher, $M_{\tilde{g}, \tilde{q}} \gg T_R$ (second scenario).  Then
for masses $m_{\tilde{G}} = 1 - 15$~keV, gravitinos can indeed serve
as warm dark matter particles. In any case, gravitino as warm dark
matter candidate will be either ruled out or supported by the LHC
experiments.

\vspace{0.5cm}

{\bf Acknowledgments.}  We are indebted to F.~Bezrukov,
  A. Boyarsky, S.~Demidov, V.~Lukash, O.~Ruchayskiy, 
  M.~Shaposhnikov  and I.~Tkachev for  the interest to this work
  and  useful
discussions.  This work was supported in part by the grants of the President of
the Russian Federation NS-1616.2008.2 and
MK-1957.2008.2 (DG), by the RFBR grant 08-02-00473-a and by the
Russian Science Support Foundation (DG).

\end{document}